\documentclass{elsart}
\usepackage{graphicx,amssymb}
\usepackage{algorithmic}
\usepackage{algorithm}
\usepackage[switch,pagewise]{lineno}
\usepackage[usenames]{color}
\usepackage{graphicx,amssymb}
\usepackage{algorithmic}
\usepackage{algorithm}
\usepackage[switch,pagewise]{lineno}
\usepackage[usenames]{color}
\usepackage{longtable}
\usepackage{amsmath}
\usepackage{threeparttable}
\usepackage{graphicx}
\usepackage{url}

\begin{document}
\begin{frontmatter}

\title{Improved Lower Bounds for Sum Coloring via Clique Decomposition}

\author{Qinghua Wu}
\ead{wu@info.univ-angers.fr} \and
\author{Jin-Kao Hao\corauthref{cor}}
\corauth[cor]{Corresponding author.} \ead{hao@info.univ-angers.fr}
\address{LERIA, Universit\'{e} d'Angers \\ 2 Boulevard Lavoisier, 49045 Angers Cedex 01, France}

\address{LERIA, Universit\'{e} d'Angers, 2 Boulevard Lavoisier, \\ 49045 Angers, Cedex 01, France}

\maketitle              
\date{June 28 2012}
\begin{abstract}

Given an undirected graph $G = (V,E)$ with a set $V$ of vertices and a set $E$ of edges, the minimum sum coloring problem (MSCP) is to find a legal vertex coloring of $G$, using colors represented by natural numbers $1, 2, . . .$ such that the total sum of the colors assigned to the vertices is minimized. This paper describes an approach based on the decomposition of the original graph into disjoint cliques for computing lower bounds for the MSCP. Basically, the proposed approach identifies and removes at each extraction iteration a maximum number of cliques of the same size (the largest possible) from the graph. Computational experiments show that this approach is able to improve on the current best lower bounds for 14 benchmark instances, and to prove optimality for the first time for 4 instances. We also report lower bounds for 24 more instances for which no such bounds are available in the literature. These new lower bounds are useful to estimate the quality of the upper bounds obtained with various heuristic approaches.

\emph{keywords}: sum coloring; graph coloring; clique decomposition; bounds; heuristics
\end{abstract}
\end{frontmatter}

\renewcommand{\baselinestretch}{1.0}\large\normalsize
\section{Introduction}

Let $G = (V,E)$ be an undirected graph with vertex set $V$ and edge set $E$. A clique is a subset of vertices $C\subseteq V$ such that each pair of vertices in $C$ are  adjacent. An independent set is a subset of vertices, $I\subseteq V$, such that no two vertices in $I$ are adjacent. A legal $k$-coloring of $G$ corresponds to a partition of $V$ into $k$ independent sets (color classes) $I_{1}$, $I_{2}$, ..., $I_{k}$. The well-known graph coloring problem aims at finding the smallest $k$ for a given graph $G$ (its chromatic number $\chi(G)$) such that $G$ has a legal $k$-coloring. In this paper, we are interested in a related problem known as the minimum sum coloring problem (MSCP for short) \cite{Kubicka1989}.

The objective of MSCP is to find a legal coloring $c=\{I_{1}$, $I_{2}$,..., $I_{k}\}$ of the graph $G$, such that the following total sum of the colors is minimized:
\begin{eqnarray}\label{sum}
Sum(c)=\sum_{i=1}^{k}\sum_{v\in I_i}^{}i
\end{eqnarray}

The optimal value for the MSCP is called the chromatic sum of $G$ and is denoted by $\sum(G)$. For any optimal solution, the associated number of required colors is termed as the strength of the graph, and denoted by $s(G)$. Clearly, the chromatic number is a lower bound of chromatic sum, i.e., $s(G) \geq \chi(G)$.

Efficient approximation algorithms or polynomial algorithms exist for specific graph classes (e.g., tree, interval graphs, line
graphs etc.) \cite{BarNoy1998,BarNoy1998B,Bonomo2011,Hajiabolhassan2000,Jansen1998,Jiang1999,Kroon1996,Kubika1991,Malafiejski2004,Salavatipour2003}. However, for the general case, the MSCP, more precisely its decision version, is known to be NP-complete \cite{Kubicka1989} and thus computationally hard. The MSCP also arises in a variety of real-world applications including those from VLSI design, scheduling and resource allocation \cite{BarNoy1998,Gandhi2004,Malafiejski2004}. 

Given the importance of the MSCP, a number of heuristics and metaheuristics have been devised to find approximate solutions (upper bounds) of good quality with a reasonable computing time. This includes greedy algorithm based on the well-known RLF graph coloring heuristic \cite{Li2009}, local search algorithms \cite{Helmar2011},  tabu search \cite{Bouziri2010}, parallel genetic algorithm \cite{Kokosinski2007}, hybrid algorithm \cite{Douiri2011} and heuristics based on independent set extraction \cite{BarNoy1998,WuHao2012}. These approaches produce upper bounds to this minimization problem. Lower bounds are useful to assess the quality of these solutions.

Recently, several studies have been devoted to determining lower bounds for the MSCP. Many of them are based on a general approach that decomposes the original graph into specific graphs like trees \cite{Kroon1996}, paths \cite{Moukrim} or cliques \cite{Douiri2012,Helmar2011,Moukrim}. Such a decomposition produces a partial graph to the original graph $G$, whose associated chromatic sum can be efficiently computed and leads naturally to a lower bound to the chromatic sum of the original graph. In \cite{Moukrim}, A. Moukrim \textit{et al.} show that the clique decomposition provides better lower bounds than other graph decompositions like trees and paths. To obtain a clique decomposition of the original graph $G$, a simple method is to determine with a vertex coloring algorithm a legal coloring of the complementary graph $\overline{G}$ of the original graph $G$. Since an independent set (i.e., a color class) of $\overline{G}$ corresponds to a clique in $G$, a coloring of $\overline{G}$ defines a clique decomposition of $G$. 

This graph coloring based approach for determining MSCP lower bounds has been exploited in studies like \cite{Douiri2012,Helmar2011,Moukrim}. These studies differ mainly by their way of coloring the complementary graph. In \cite{Moukrim}, the coloring is achieved with an adjusted greedy algorithm MRLF, while an ant colony optimization algorithm is used in \cite{Douiri2012} and a local search heuristic is employed in \cite{Helmar2011} to color $\overline{G}$. Finally, some theoretical lower bounds for the MSCP are proposed in \cite{Kokosinski2007,Kubicka2004}.

In this paper, we present a heuristic approach (denoted by EXCLIQUE) based on a direct clique decomposition of $G$ for computing the lower bounds for the general MSCP. The proposed approach tries to extract as many large cliques as possible from the graph. Since more colors and larger color numbers are needed to color the vertices of a large clique than to color the vertices of a small clique, a clique decomposition with more and large cliques tends to increase the chromatic sum of the resulting clique decomposition and therefore leads to a tighter lower bound to the original graph. To achieve this, we follow a similar approach which has been successfully used to establish \textit{upper bounds} for the vertex coloring problem \cite{WuHaoCOR} and the MSCP \cite{WuHao2012}. At each iteration of the proposed approach, a maximum number of pairwise disjoint cliques of the largest possible size are identified and removed from the graph. This process is repeated until the graph becomes empty. The proposed approach is assessed on a set of 62 DIMACS and COLOR2 benchmark graphs in the literature. The computational outcomes show that this approach is able to improve on the current best lower bounds in 14 cases out of 38 instances with known lower bounds reported in the literature, and to prove optimality for the first time for 4 instances. Lower bounds are also reported for the first time for the remaining 24 instances for which no lower bounds are available in the literature.



The rest of the paper is organized as follows. In section 2, we explain the clique decomposition approach for computing lower bounds for the MSCP. In section 3, our proposed heuristic approach is presented. In section 4, we provide computational results and comparisons on a set of 62 benchmark graphs from the literature. In section 5, we show additional studies about the effect of the disjoint clique removal strategy, followed by conclusions in section 6.

\section{Lower bounds for the MSCP based on clique decomposition}

Given $G=(V,E)$, a partial graph of $G=(V,E)$ is a graph $G'=(V,E')$ such that $E'$ is a subset of $E$. It is easy to observe that the chromatic sum of $G'$ is a lower bound for the chromatic sum of $G$. Indeed, any legal coloring of $G$ is a legal coloring of $G'$ while the reverse does not hold. Thus, to calculate a lower bound for the chromatic sum of the original graph $G$, one could try to find a partial graph of the original graph whose chromatic sum can be efficiently computed and maximized.

\begin{figure}[!hbp]
  \centering\scalebox{0.70}{\includegraphics{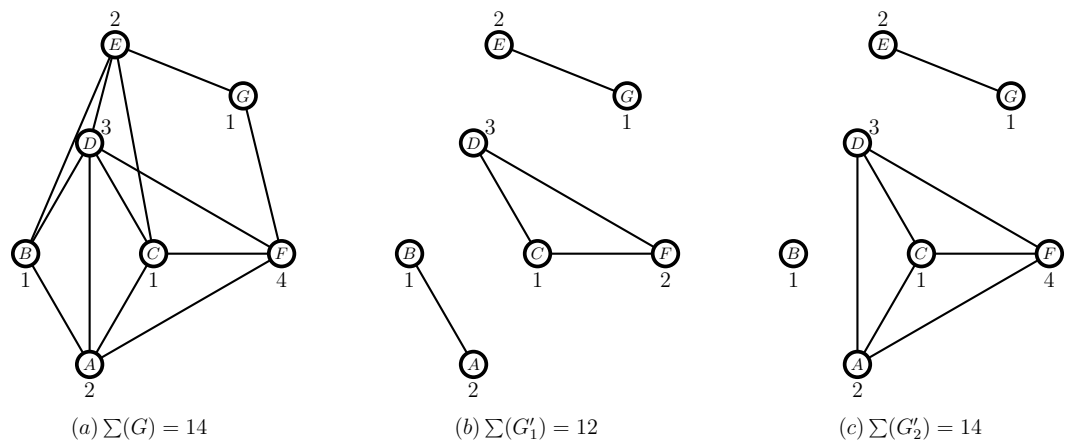}}\\
  \centering\caption{Partial graphs of $G$ via clique decomposition}
  \label{lowbound}
\end{figure}

This can be achieved by decomposing the vertex set of $G$ into $k$ pairwise disjoint cliques $C_{1}$, $C_{2}$ ,..., $C_{p}$ such that $\forall i \neq j, \  C_{i}\bigcap C_{j}=\emptyset$ and $\bigcup_{i}C_{i}=V$ (see Fig.\ref{lowbound} for an illustrative example). Given a clique decomposition $\digamma=\{C_{1}$, $C_{2}$,..., $C_{p}\}$, each of its cliques $C_i$ ($i=1..p)$ needs exactly $|C_{i}|$ colors: $1..|C_{i}|$. Consequently, the chromatic sum of $\digamma$ is given by $\sum_{i=1}^{p}\frac{|C_{i}|(|C_{i}|+1)}{2}$. Since a clique decomposition $\digamma=\{C_{1}$, $C_{2}$,..., $C_{p}\}$ is a partial graph of $G$ and the chromatic sum of $\digamma$ is therefore a lower bound to the chromatic sum $\sum(G)$ of $G$. 

The quality of this lower bound depends on the way to decompose the graph into cliques. For instance, consider the graph $G=(V,E)$ of Fig. \ref{lowbound}(a), we decompose $G$ in two different ways (Fig. \ref{lowbound}(b) and Fig. \ref{lowbound}(c)), we obtain the following chromatic sum $\sum(G_{1}')=12$ and $\sum(G_{2}')=14$. Thus, in order to obtain a lower bound as tight (large) as possible, one could try to find a clique decomposition of $G$ whose chromatic sum is as large as possible. This can be considered as an optimization problem where we search for a decomposition of $G$ into cliques, such that the associated chromatic sum is maximized over all the possible clique decompositions:
\begin{equation}\label{lbbest}
LB^{*}=max\{ \ \sum(\digamma) \ |  \ \digamma \ is \ a \ clique \ decomposition \ of \ G  \ \}
\end{equation}

It should be noted that $LB^{*}$ might be strictly smaller than $\sum(G)$ because, a clique decomposition, by ignoring some edges of the original graph $G$, is a less constrained problem for coloring.

This clique decomposition approach for computing lower bounds for the MSCP was originally proposed in \cite{Moukrim} and were further exploited in \cite{Douiri2012,Helmar2011}. In these previous studies, clique decompositions are obtained by apply a vertex coloring algorithm to color the complementary graph $\overline{G}$ and to use the color classes to define a clique decomposition of $G$. 


In this work, we present a heuristic approach (denoted by EXCLIQUE) based on a direct clique decomposition of the original graph $G$ to approximate $LB^{*}$. In order to obtain a clique decomposition which maximizes its chromatic sum, the proposed approach tries to extract as many large cliques as possible from the original graph by employing a recent maximum clique algorithm presented in \cite{WuHao2010a}. 

\section{EXCLIQUE: an algorithm for improved lower bounds of MSCP}

\subsection{Rationale and general procedure}

Given a clique decomposition $\digamma=\{C_{1}$, $C_{2}$,..., $C_{k}\}$ of $G=(V,E)$, there is one unique coloring for $\digamma$, i.e., for each clique $C_{i}$ ($1\leq i \leq k$) of $\digamma$, we need exactly $|C_{i}|$ colors to color its vertices. Moreover, the larger the clique $C_{i}$ is, the more the number of vertices in $C_{i}$ that need to be colored with large colors. Thus, a clique decomposition with more large cliques tends to have a larger chromatic sum and therefore gives a better lower bound to $G$. For instance, consider again the example of Fig. \ref{lowbound}. Vertex A belongs respectively to a clique of size 2 (Fig.\ref{lowbound}(b)) and a clique of size 4 (Fig.\ref{lowbound}(c)). The later case increases the sum of colors by 2 because we need a large color (4) for the clique of size 4. 

For the purpose of obtaining a clique decomposition with more large cliques, one could try to identify as many large cliques as possible from the graph. To achieve this, we can iteratively identify and remove the maximum number of disjoint cliques of the maximum size from the graph until the graph becomes empty.
Our proposed EXCLIQUE algorithm follows the above idea and can be summarized as follows:
\begin{enumerate}
  \item Identify a maximum clique $C$ in $G$;
  \item Collect in a set $M$ as many other cliques of size $|C|$ as possible;
  \item Find from $M$ a maximum number of \textit{disjoint} cliques $C_{1}$, ..., $C_{t}$;
  \item Remove $C_{1}$, ..., $C_{t}$ from $G$;
  \item Repeat the above steps until the graph becomes empty;
\end{enumerate}

In step 1 of the above process, one needs to identify a maximum clique. Notice that finding a maximum clique in a graph is an NP-hard problem in the general case \cite{Garey1979}. For this reason, we use the so-called adaptive tabu search heuristic (denoted by ATS) designed for the maximum clique problem to find large cliques \cite{WuHao2010a}. The same heuristic is also employed to build the pool $M$ composed of cliques of a given size (step 2) and identify the set of disjoint cliques (step 3).

In step 2, in order to build the pool $M$ of cliques of size $|C|$, we apply repeatedly ATS to generate as many cliques of size $|C|$ as possible and put them in set $M$. The search for a new clique of size $|C|$ stops when the number of cliques contained in $M$ reaches a desired threshold ($M_{max}$) or when no new clique of size $|C|$ is found after $p_{max}$ consecutive tries.

In step 3, it is required to find among the candidates of $M$ a maximum number of disjoint cliques. This task corresponds in fact to the \textit{maximum set packing} problem, which is equivalent to the maximum clique problem \cite{Garey1979}. To see this, it suffices to construct an instance of the maximum clique problem from $M = \{C_{1}, ..., C_{n}\}$ as follows. Define a new graph $G'=(V',E')$ where $V'=\{1,...,n\}$ and $\{i,j\} \in E'$ $(i, j \in V')$ if and only if $C_{i}$ and $C_{j}$ share no common element, i.e., $C_{i}\cap C_{j} = \emptyset$. Now it is clear that there is a strict equivalence between a clique in $G'=(V',E')$ and a set of disjoint cliques in $M$. Consequently, one can apply again the ATS algorithm to approximate this problem.

\begin{figure}[!hbp]
  \centering\scalebox{0.70}{\includegraphics{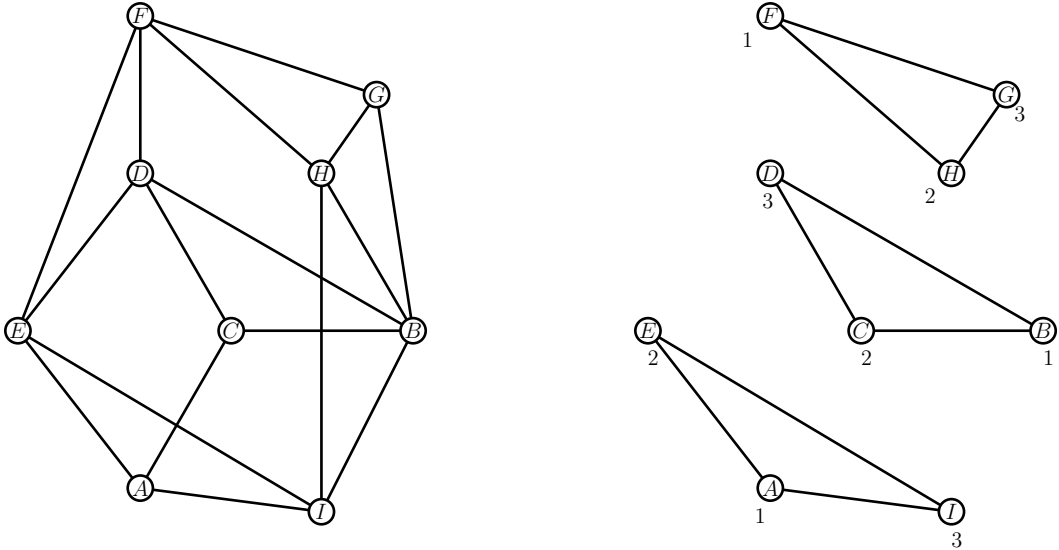}}\\
  \centering\caption{An illustration of the proposed EXCLIQUE algorithm}
  \label{clique}
\end{figure}

Fig. \ref{clique} illustrates how our EXCLIQUE approach works on a graph with 9 vertices. At the first step, we find a maximum clique of size 3 from the graph (e.g., $\{D,E,F\}$). Then
we try to collect as many cliques of size 3 as possible in $M$, leading to a pool containing 6 cliques of size 3: $\{A,E,I\}$, $\{B,C,D\}$, $\{B,G,H\}$, $\{D,E,F\}$, $\{F,H,G\}$ and $\{G,F,H\}$. From these cliques, a set of 3 disjoint cliques $\{A,E,I\}$, $\{B,C,D\}$, $\{G,F,H\}$ is identified. Since the graph becomes empty after removing these 3 cliques, the procedure stops. We obtain a clique decomposition $\digamma=\{\{A,E,I\},\{B,C,D\},\{G,F,H\}\}$ with a chromatic sum value equal to 18. 

Note that if the first clique $\{D,E,F\}$ would have been removed once it was identified, then all the resulting clique decompositions would have a chromatic sum inferior to 18, leading to a worse lower bound for graph $G$.

\section{Experimental results}

To assess the efficiency of the proposed EXCLIQUE approach, we conduct experiments on two sets of benchmark instances from the literature. The first set is composed of 29 well-known DIMACS graphs\footnote{\url{ftp://dimacs.rutgers.edu/pub/challenge/graph/benchmarks/color/}}, which are very popular for testing graph coloring algorithms \cite{Johnson1996}. The second set of benchmarks is composed of 33 graphs from the the COLOR02 competition\footnote{\url{http://mat.gsia.cmu.edu/COLOR02/}}. In addition, we also assess the interest of the basic extraction method which extracts maximum cliques one by one and the pertinence of applying directly powerful graph vertex coloring algorithms on the complementary graph of the original graph.

Our algorithm is implemented in C++, and compiled with GNU gcc on an Intel Xeon E5440 processor with 2.83 GHz and 8GB RAM. To obtain our
computational results, each instance is solved 20 times independently with different random seeds (the two very large instances Cxxxx.5 are solved 5 times). EXCLIQUE stops when the graph under consideration becomes empty.

The two main parameters for EXCLIQUE are $M_{max}$ and $p_{max}$. Obviously, larger values for $M_{max}$ and $p_{max}$ could include more cliques in $M$, thus giving a higher chance of finding more pairwise disjoint clique sets of the largest possible size in $M$. On the other hand, larger values for both parameters also imply longer computing time. According to our experiments, we have fixed $p_{max} = 100$ and $M_{max} = 2000$ for all our experiments. In addition to $M_{max}$ and $p_{max}$, ATS requires also several parameters. In our case, we adopt those used in
the original paper \cite{WuHao2010a}.

\subsection{Improved lower bounds for the MSCP}

Tables \ref{dimacs} and \ref{color02} respectively show the computational statistics of the EXCLIQUE algorithm on the DIMACS benchmarks and the COLOR02 instances. In both tables, columns 2--4 indicate the features of the tested graphs, including the number of vertices ($|V|$), the number of edges ($|E|$) and the density of the graph ($Den$). Columns 5 and 6 give respectively the current best upper and lower bounds reported in the literature \cite{Bouziri2010,Douiri2011,Douiri2012,Helmar2011,Kokosinski2007,Li2009,Moukrim,WuHao2012}. The results of our EXCLIQUE approach are given in columns 7–-9, including the best lower bounds found by our EXCLIQUE approach over the 20 runs, the averaged lower bound value with the standard deviation between parentheses and the average CPU time in seconds.

\renewcommand{\baselinestretch}{0.7}\large\normalsize

\begin{table}\centering
\begin{small}
\caption{Computational results of EXCLIQUE on 29 DIMACS challenge benchmarks. The symbol '-' means that the information is not available. Improved bounds or new bounds are indicated in bold.} \label{dimacs}
\begin{tabular}{p{2.0cm}|p{0.7cm}p{1.0cm}p{0.7cm}p{1.0cm}p{0.7cm}p{1.3cm}p{3.0cm}p{1.20cm}p{0cm}}
\hline
\leftline{$Instance$} &  \centering{$|V|$} & \centering{$|E|$} & \centering{$Den$} & \centering{$UB^{*}$} & \centering{$LB^{*}$} & \multicolumn{3}{c}{EXCLIQUE} &\\
\cline{7-10}
& & &  & &  & \centering{$LB$} & \centering{$Avg.(Std.)$} &  \centering{$T[second]$} &  \\
\hline
\centering{DSJC125.1} & \centering{125} & \centering{736} & \centering{0.09} & \centering{326} & \centering{238} & \centering{\textbf{246}} & \centering{244.10(0.94)} & \centering{$80$} &  \\
\centering{DSJC125.5} & \centering{125} & \centering{3891} & \centering{0.50} & \centering{1015} & \centering{504} & \centering{\textbf{536}} & \centering{522.40(5.46)} & \centering{$35$} &  \\
\centering{DSJC125.9} & \centering{125} & \centering{6961} & \centering{0.89} & \centering{2511} & \centering{1621} & \centering{\textbf{1664}} & \centering{1592.50(26.61)} & \centering{$47$} & \\
\centering{DSJC250.1} & \centering{250} & \centering{3218} & \centering{0.10} & \centering{977} & \centering{537} & \centering{\textbf{567}} & \centering{561.95(1.96)} & \centering{$46$} & \\
\centering{DSJC250.5} & \centering{250} & \centering{15668} & \centering{0.50} & \centering{3246} & \centering{1150} & \centering{\textbf{1270}} & \centering{1258.80(5.44)} & \centering{$37$} & \\
\centering{DSJC250.9} & \centering{250} & \centering{27897} & \centering{0.90} & \centering{8286} & \centering{3972} & \centering{\textbf{4179}} & \centering{4082.40(51.66)} & \centering{$158$} & \\
\centering{DSJC500.1} & \centering{500} & \centering{12458} & \centering{0.10} & \centering{2850} & \centering{1163} & \centering{\textbf{1250}} & \centering{1246.55(1.37)} & \centering{$1269$} & \\
\centering{DSJC500.5} & \centering{500} & \centering{62624} & \centering{0.50} & \centering{10910} & \centering{2616} & \centering{\textbf{2921}} & \centering{2902.60(11.94)} & \centering{$60$} & \\
\centering{DSJC500.9} & \centering{500} & \centering{112437} & \centering{0.90} & \centering{29912} & \centering{10074} & \centering{\textbf{10881}} & \centering{10734.50(74.30)} & \centering{$276$} & \\
\centering{DSJC1000.1} & \centering{1000} & \centering{49629} & \centering{0.10} & \centering{9003} & \centering{2499} & \centering{\textbf{2762}} & \centering{2758.55(2.13)} & \centering{$5193$} & \\
\centering{DSJC1000.5} & \centering{1000} & \centering{249826} & \centering{0.50} & \centering{37598} & \centering{5787} & \centering{\textbf{6708}} & \centering{6665.90(14.49)} & \centering{$155$} & \\
\centering{DSJC1000.9} & \centering{1000} & \centering{449449} & \centering{0.90} & \centering{103464} & \centering{23863} & \centering{\textbf{26557}} & \centering{26300.25(84.04)} & \centering{$2741$} & \\
\centering{flat300\_20\_0} & \centering{300} & \centering{21375} & \centering{0.48} & \centering{3150} & \centering{-} & \centering{\textbf{1524}} & \centering{1505.65(6.78)} & \centering{$35$} & \\
\centering{flat300\_26\_0} & \centering{300} & \centering{21633} & \centering{0.48} & \centering{3966} & \centering{-} & \centering{\textbf{1525}} & \centering{1511.40(8.40)} & \centering{$34$} & \\
\centering{flat300\_28\_0} & \centering{300} & \centering{21695} & \centering{0.48} & \centering{4282} & \centering{-} & \centering{\textbf{1532}} & \centering{1515.25(7.81)} & \centering{$43$} & \\
\centering{flat1000\_50\_0} & \centering{1000} & \centering{245000} & \centering{0.49} & \centering{25500} & \centering{-} & \centering{\textbf{6601}} & \centering{6571.80(15.54)} & \centering{$118$} & \\
\centering{flat1000\_60\_0} & \centering{1000} & \centering{245830} & \centering{0.49} & \centering{30100} & \centering{-} & \centering{\textbf{6640}} & \centering{6600.50(18.01)} & \centering{$414$} & \\
\centering{flat1000\_76\_0} & \centering{1000} & \centering{246708} & \centering{0.49} & \centering{37167} & \centering{-} & \centering{\textbf{6632}} & \centering{6583.15(17.53)} & \centering{$98$} & \\
\centering{le450\_15a} & \centering{450} & \centering{8168} & \centering{0.08} & \centering{2632} & \centering{-} & \centering{\textbf{2329}} & \centering{2313.65(15.32)} & \centering{$252$} & \\
\centering{le450\_15b} & \centering{450} & \centering{8169} & \centering{0.08} & \centering{2642} & \centering{-} & \centering{\textbf{2343}} & \centering{2315.65(15.05)} & \centering{$600$} & \\
\centering{le450\_15c} & \centering{450} & \centering{16680} & \centering{0.17} & \centering{3866} & \centering{-} & \centering{\textbf{2591}} & \centering{2545.30(24.67)} & \centering{$187$} & \\
\centering{le450\_15d} & \centering{450} & \centering{16750} & \centering{0.17} & \centering{3921} & \centering{-} & \centering{\textbf{2610}} & \centering{2572.40(24.13)} & \centering{$175$} & \\
\centering{le450\_25a} & \centering{450} & \centering{8260} & \centering{0.08} & \centering{3153} & \centering{-} & \centering{\textbf{2997}} & \centering{2964.40(28.08)} & \centering{$967$} &\\
\centering{le450\_25b} & \centering{450} & \centering{8263} & \centering{0.08} & \centering{3366} & \centering{-} & \centering{\textbf{3305}} & \centering{3304.10(0.70)} & \centering{$1550$} & \\
\centering{le450\_25c} & \centering{450} & \centering{17343} & \centering{0.17} & \centering{4515} & \centering{-} & \centering{\textbf{3619}} & \centering{3597.10(11.82)} & \centering{$689$} & \\
\centering{le450\_25d} & \centering{450} & \centering{17425} & \centering{0.17} & \centering{4544} & \centering{-} & \centering{\textbf{3684}} & \centering{3627.35(45.33)} & \centering{$850$} & \\
\centering{latin\_sqr\_10} & \centering{900} & \centering{307350} & \centering{0.76} & \centering{42223} & \centering{-} & \centering{\textbf{40950}} & \centering{40950(0.00)} & \centering{$15$} & \\
\centering{C2000.5} & \centering{2000} & \centering{999836} & \centering{0.50} & \centering{132515} & \centering{-} & \centering{\textbf{15091}} & \centering{15077.60(11.74)} & \centering{$3994$} & \\
\centering{C4000.5} & \centering{4000} & \centering{4000268} & \centering{0.50} & \centering{473234} & \centering{-} & \centering{\textbf{33033}} & \centering{33018.80(11.42)} & \centering{$14413$} & \\
\hline
\end{tabular}

\end{small}
\end{table}

\renewcommand{\baselinestretch}{0.7}\large\normalsize

\begin{table}\centering
\begin{small}
\caption{Computational results of EXCLIQUE on 33 COLOR02 challenge benchmarks. The symbol '-' means that the information is not available. Improved bounds or new bounds are indicated in bold.} \label{color02}
\begin{tabular}{p{2.4cm}|p{0.7cm}p{1.0cm}p{0.7cm}p{1.0cm}p{0.7cm}p{1.3cm}p{3.0cm}p{1.20cm}p{0cm}}
\hline
\leftline{$Instance$} &  \centering{$|V|$} & \centering{$|E|$} & \centering{$Den$} & \centering{$UB^{*}$} & \centering{$LB^{*}$} & \multicolumn{3}{c}{EXSCOL} &\\
\cline{7-10}
& & &  & &  & \centering{$LB$} & \centering{$Avg.(Std.)$} &  \centering{$T[second]$} &  \\
\hline
\centering{myciel3} & \centering{11} & \centering{20} & \centering{0.40} & \centering{21} & \centering{16} & \centering{16} & \centering{16(0.00)} & \centering{$25$} & \\
\centering{myciel4} & \centering{23} & \centering{71} & \centering{0.28} & \centering{45} & \centering{34} & \centering{34} & \centering{34(0.00)} & \centering{$42$} & \\
\centering{myciel5} & \centering{47} & \centering{236} & \centering{0.22} & \centering{93} & \centering{70} & \centering{70} & \centering{70(0.00)} & \centering{$73$} &\\
\centering{myciel6} & \centering{95} & \centering{755} & \centering{0.17} & \centering{189} & \centering{142} & \centering{142} & \centering{142(0.00)} & \centering{$93$} & \\
\centering{myciel7} & \centering{191} & \centering{2360} & \centering{0.13} & \centering{381} & \centering{286} & \centering{286} & \centering{286(0.00)} & \centering{$148$} & \\
\centering{anna} & \centering{138} & \centering{493} & \centering{0.05} & \centering{277} & \centering{273} & \centering{273} & \centering{273(0.00)} & \centering{$168$} & \\
\centering{david} & \centering{87} & \centering{406} & \centering{0.11} & \centering{241} & \centering{234} & \centering{\emph{229}} & \centering{229(0.00)} & \centering{$73$} & \\
\centering{huck} & \centering{74} & \centering{301} & \centering{0.11} & \centering{243} & \centering{243} & \centering{243} & \centering{243(0.00)} & \centering{$58$} & \\
\centering{jean} & \centering{80} & \centering{254} & \centering{0.08} & \centering{217} & \centering{216} & \centering{216} & \centering{216(0.00)} & \centering{$67$} & \\
\centering{queen5.5} & \centering{25} & \centering{160} & \centering{0.53} & \centering{75} & \centering{75} & \centering{75} & \centering{75(0.00)} & \centering{$19$} & \\
\centering{queen6.6} & \centering{36} & \centering{290} & \centering{0.46} & \centering{138} & \centering{126} & \centering{126} & \centering{126(0.00)} & \centering{$28$} & \\
\centering{queen7.7} & \centering{49} & \centering{476} & \centering{0.40} & \centering{196} & \centering{196} & \centering{196} & \centering{196(0.00)} & \centering{$41$} & \\
\centering{queen8.8} & \centering{64} & \centering{728} & \centering{0.36} & \centering{302} & \centering{288} & \centering{288} & \centering{288(0.00)} & \centering{$66$} & \\
\centering{games120} & \centering{120} & \centering{638} & \centering{0.09} & \centering{446} & \centering{442} & \centering{442} & \centering{441.40(0.91)} & \centering{$105$} & \\
\centering{miles250} & \centering{128} & \centering{387} & \centering{0.05} & \centering{334} & \centering{318} & \centering{318} & \centering{316.15(0.35)} & \centering{$131$} & \\
\centering{miles500} & \centering{128} & \centering{1170} & \centering{0.14} & \centering{715} & \centering{686} & \centering{\emph{677}} & \centering{671.35(3.27)} & \centering{$117$} &\\
\centering{wap05} & \centering{905} & \centering{43081} & \centering{0.10} & \centering{13680} & \centering{-} & \centering{\textbf{12428}} & \centering{12339.25(44.03)} & \centering{$6283$} & \\
\centering{wap06} & \centering{947} & \centering{43571} & \centering{0.10} & \centering{13778} & \centering{-} & \centering{\textbf{12393}} & \centering{12348.75(43.59)} & \centering{$5417$} & \\
\centering{wap07} & \centering{1809} & \centering{103368} & \centering{0.06} & \centering{28629} & \centering{-} & \centering{\textbf{24339}} & \centering{24263.82(52.18)} & \centering{$8359$} & \\
\centering{wap08} & \centering{1870} & \centering{104176} & \centering{0.06} & \centering{28896} & \centering{-} & \centering{\textbf{24791}} & \centering{24681.09(56.12)} & \centering{$9127$} & \\
\centering{qg.order30} & \centering{900} & \centering{26100} & \centering{0.06} & \centering{\underline{13950}} & \centering{-} & \centering{\textbf{\underline{13950}}} & \centering{13950(0.00)} & \centering{474} & \\
\centering{qg.order40} & \centering{1600} & \centering{62400} & \centering{0.05} & \centering{\underline{32800}} & \centering{-} & \centering{\underline{\textbf{32800}}} & \centering{32800(0.00)} & \centering{$1379$} & \\
\centering{qg.order60} & \centering{3600} & \centering{212400} & \centering{0.03} & \centering{110925} & \centering{-} & \centering{\textbf{109800}} & \centering{109800(0.00)} & \centering{$7507$} & \\
\centering{2-Insertions\_3} & \centering{37} & \centering{72} & \centering{0.11} & \centering{62} & \centering{55} & \centering{55} & \centering{55(0.00)} & \centering{9} & \\
\centering{3-Insertions\_3} & \centering{56} & \centering{110} & \centering{0.07} & \centering{92} & \centering{84} & \centering{84} & \centering{82.8(0.50)} & \centering{12} & \\
\centering{fpsol2.i.1} & \centering{496} & \centering{11654} & \centering{0.09} & \centering{\underline{3403}} & \centering{3402} & \centering{\underline{\textbf{3403}}} & \centering{3403(0.00)} & \centering{2676} & \\
\centering{inithx.i.1} & \centering{864} & \centering{18707} & \centering{0.05} & \centering{\underline{3676}} & \centering{3581} & \centering{\underline{\textbf{3676}}} & \centering{3676(0.00)} & \centering{3689} & \\
\centering{mug100\_1} & \centering{100} & \centering{166} & \centering{0.03} & \centering{202} & \centering{188} & \centering{188} & \centering{188(0.00)} & \centering{24} & \\
\centering{mug100\_25} & \centering{100} & \centering{166} & \centering{0.03} & \centering{202} & \centering{186} & \centering{186} & \centering{183.35(0.57)} & \centering{36} & \\
\centering{mug88\_1} & \centering{88} & \centering{146} & \centering{0.04} & \centering{178} & \centering{164} & \centering{164} & \centering{162.25(0.34)} & \centering{27} & \\
\centering{mug88\_25} & \centering{88} & \centering{146} & \centering{0.04} & \centering{178} & \centering{162} & \centering{162} & \centering{160.25(0.43)} & \centering{29} & \\
\centering{zeroin.i.2} & \centering{211} & \centering{3541} & \centering{0.16} & \centering{1004} & \centering{1004} & \centering{1004} & \centering{1004(0.00)} & \centering{453} & \\
\centering{zeroin.i.3} & \centering{206} & \centering{3540} & \centering{0.17} & \centering{998} & \centering{998} & \centering{998} & \centering{998(0.00)} & \centering{442} & \\

\hline
\end{tabular}

\end{small}
\end{table}

\renewcommand{\baselinestretch}{1.0}\large\normalsize

Concerning the 29 DIMACS instances, from Table \ref{dimacs}, we observe that the results obtained by EXCLIQUE are very competitive when compared to the current best lower bounds reported in the literature. Indeed, for the 12 random DSJC graphs with a known lower bound, we managed to improve on the current best bounds in all the cases. For the remaining 17 DIMACS graphs, we report the computational statistics for the lower bounds for the first time. Finally, Table \ref{dimacs} also discloses that for most of these DIMACS instances, the gaps between the best known upper bounds and our best lower bounds are still large.

Concerning the 33 COLOR02 instances, from Table \ref{color02}, we observe that our EXCLIQUE algorithm is able to improve the current best known lower bounds in the literature for 2 instances (fpsol2.i.1 and inithx.i.1) while equaling the best known lower bounds for 22 instances. Only for 2 instances (david and miles500), EXCLIQUE obtains a worse result. For the other 7 COLOR02 instances, we report for the first time lower bounds. Table \ref{color02}  also indicates that we are able to prove optimality for 10 instances. Among these 10 instances, the optimality of 4 instances (qg.order30, qg.order40, fpsol2.i.1 and inithx.i.1) is proven for the first time.

\subsection{Comparison with three state-of-the-art approaches}

In this section, we compare EXCLIQUE with 3 recent state of the art algorithms for computing lower bounds for the MSCP from the literature: MDS(5)+LS \cite{Helmar2011}, RMDS(n) \cite{Moukrim} and ANT \cite{Douiri2012}. All these reference algorithms are based on the clique decomposition method which consists in transforming the original graph to its complement and then coloring the complementary graph with different coloring algorithms. First, we recall the basic experimental conditions (when they are available) used by these reference methods.

\begin{itemize}
  \item MDS(5)+LS. The results of MDS(5)+LS were based on an Intel Core i7 processor 2.93Ghz with 8192MB cache L2 and 4 GB RAM. For each instance, MDS(5)+LS was run only 1 time with a time limit of 1 hour.
  \item RMDS(n). RMDS(n) was run on an Intel Core 2 Duo T5450 1.66-1.67 with 2GB Ram. For each instance, RMDS(n) repeats the MRLF algorithm \cite{Li2009} $n$ times ($n=|V|$) starting from a different vertex for each replicate.
  \item ANT. ANT was run on an AMD Athlon(tm)X2 dual-core QL-65 (2cpus) 2.1GHz PC with 4GB RAM. For each instance, 20 runs of ANT were performed.
\end{itemize}

In Table \ref{comp}, we show the best lower bounds obtained by EXCLIQUE in comparison with the other three approaches on 38 benchmark instances (results of the reference algorithms are not available for other instances). The results of MDS(5)+LS and RMDS(n) are directly extracted from \cite{Helmar2011}. From Table \ref{comp}, we observe that our EXCLIQUE algorithm competes favorably with these 3 algorithms. Indeed, in comparison with each of these 3 algorithms, EXCLIQUE is able to achieve tighter lower bounds for  at least 9 instances (indicated in bold) and only for at most 2 instances, EXCLIQUE's result is worse than that of these reference algorithms (indicated in italics). These comparative results confirm the effectiveness of the proposed approach to deliver improved lower bounds for the MSCP.

\renewcommand{\baselinestretch}{0.7}\large\normalsize
\begin{table}\centering
\begin{small}
\caption{Comparison of lower bounds on the 38 instances reported in \cite{Helmar2011}. The symbol '-' means that the related statistics are not available. The best $LB$ values are highlighted in bold.}
\label{comp}
\begin{tabular}{p{2.5cm}|p{1.5cm}p{2.0cm}p{2.5cm}p{2.0cm}p{2.0cm}p{0cm}}
\hline
\leftline{$Instance$} & \centering{$LB^{*}$} &  \multicolumn{4}{c}{Lower bounds} &\\
\cline{3-6}
& & \centering{EXCLIQUE} &  \centering{MDS(5)+LS\cite{Helmar2011}} & \centering{RMDS(n)\cite{Moukrim}} & \centering{ANT\cite{Douiri2012}} & \\
& &  &  \centering{(2011)} & \centering{(2010)} & \centering{(2012)} & \\
\hline
\centering{DSJC125.1} & \centering{238}  & \centering{\textbf{246}} &  \centering{238} & \centering{238} & \centering{-} & \\
\centering{DSJC125.5} & \centering{504}  & \centering{\textbf{536}} &  \centering{493} & \centering{504} & \centering{-} & \\
\centering{DSJC125.9} & \centering{1621}  & \centering{\textbf{1664}} &  \centering{1621} & \centering{1600} & \centering{-} & \\
\centering{DSJC250.1} & \centering{537}  & \centering{\textbf{567}} &  \centering{521} & \centering{537} & \centering{-} & \\
\centering{DSJC250.5} & \centering{1150}  & \centering{\textbf{1270}} &  \centering{1128} & \centering{1150} & \centering{-} & \\
\centering{DSJC250.9} & \centering{3972}  & \centering{\textbf{4179}} &  \centering{3779} & \centering{3972} & \centering{-} & \\
\centering{DSJC500.1} & \centering{1163}  & \centering{\textbf{1250}} &  \centering{1143} & \centering{1163} & \centering{-} & \\
\centering{DSJC500.5} & \centering{2616}  & \centering{\textbf{2921}} &  \centering{2565} & \centering{2616} & \centering{-} & \\
\centering{DSJC500.9} & \centering{10074}  & \centering{\textbf{10881}} &  \centering{9731} & \centering{10074} & \centering{-} & \\
\centering{DSJC1000.1} & \centering{2499}  & \centering{\textbf{2762}} &  \centering{2456} & \centering{2499} & \centering{-} & \\
\centering{DSJC1000.5} & \centering{5787}  & \centering{\textbf{6708}} &  \centering{5660} & \centering{5787} & \centering{-} & \\
\centering{DSJC1000.9} & \centering{23863}  & \centering{\textbf{26557}} &  \centering{23208} & \centering{23863} & \centering{-} & \\
\centering{myciel3} & \centering{16}  & \centering{16} &  \centering{16} & \centering{16} & \centering{16} & \\
\centering{myciel4} & \centering{34}  & \centering{34} &  \centering{34} & \centering{34} & \centering{34} & \\
\centering{myciel5} & \centering{70}  & \centering{70} &  \centering{70} & \centering{70} & \centering{70} & \\
\centering{myciel6} & \centering{142}  & \centering{142} &  \centering{142} & \centering{142} & \centering{142} & \\
\centering{myciel7} & \centering{286}  & \centering{286} &  \centering{286} & \centering{286} & \centering{286} & \\
\centering{anna} & \centering{273}  & \centering{273} &  \centering{273} & \centering{272} & \centering{272} & \\
\centering{david} & \centering{234}  & \centering{\emph{229}} &  \centering{\textbf{234}} & \centering{\textbf{234}} & \centering{\textbf{234}} & \\
\centering{huck} & \centering{243}  & \centering{243} &  \centering{243} & \centering{243} & \centering{243} & \\
\centering{jean} & \centering{216}  & \centering{216} &  \centering{216} & \centering{216} & \centering{216} & \\
\centering{queen5.5} & \centering{75}  & \centering{75} &  \centering{75} & \centering{75} & \centering{75} & \\
\centering{queen6.6} & \centering{126}  & \centering{126} &  \centering{126} & \centering{126} & \centering{126} & \\
\centering{queen7.7} & \centering{196}  & \centering{196} &  \centering{196} & \centering{196} & \centering{196} & \\
\centering{queen8.8} & \centering{288}  & \centering{288} &  \centering{288} & \centering{288} & \centering{288} & \\
\centering{games120} & \centering{442}  & \centering{442} &  \centering{442} & \centering{442} & \centering{442} & \\
\centering{miles250} & \centering{318}  & \centering{318} &  \centering{318} & \centering{316} & \centering{316} & \\
\centering{miles500} & \centering{686}  & \centering{\emph{677}} &  \centering{\textbf{686}} & \centering{677} & \centering{677} & \\
\centering{2-Insertions\_3} & \centering{55}  & \centering{55} &  \centering{55} & \centering{55} & \centering{55} & \\
\centering{3-Insertions\_3} & \centering{84}  & \centering{84} &  \centering{84} & \centering{84} & \centering{84} & \\
\centering{fpsol2.i.1} & \centering{3402}  & \centering{\textbf{3403}} &  \centering{3151} & \centering{3402} & \centering{2590} & \\
\centering{inithx.i.1} & \centering{3581}  & \centering{\textbf{3676}} &  \centering{3486} & \centering{3581} & \centering{2801} & \\
\centering{mug88\_1} & \centering{188}  & \centering{188} &  \centering{188} & \centering{186} & \centering{187} & \\
\centering{mug88\_25} & \centering{186}  & \centering{186} &  \centering{186} & \centering{183} & \centering{185} & \\
\centering{mug100\_1} & \centering{164}  & \centering{164} &  \centering{164} & \centering{163} & \centering{163} & \\
\centering{mug100\_25} & \centering{162}  & \centering{162} &  \centering{162} & \centering{161} & \centering{162} & \\
\centering{zeroin.i.2} & \centering{1004}  & \centering{1004} &  \centering{1004} & \centering{1004} & \centering{1003} & \\
\centering{zeroin.i.3} & \centering{998}  & \centering{998} &  \centering{998} & \centering{998} & \centering{997} & \\
\hline
\end{tabular}

\end{small}
\end{table}

\renewcommand{\baselinestretch}{1.0}\large\normalsize

\subsection{Comparison with the graph coloring approach}


Using graph coloring methods to compute the lower bound of the MSCP is exploited in \cite{Douiri2012,Helmar2011,Moukrim}. With this approach, a graph coloring algorithm is applied to the complement of the original graph $G$ and the color classes of the coloring induces a clique decomposition of $G$. In this section, we revisit this approach by employing one of the most recent and effective coloring algorithm (i.e., Memetic Coloring Algorithm--MACOL \cite{lu2010}) and show a comparison with the EXCLIQUE approach. 



For this purpose, we consider the 12 DSJC random graphs and run MACOL 20 times to solve each instance. The results of EXCLIQUE are extracted from Table \ref{dimacs}. The comparative results between MACOL and EXCLIQUE are summarized in Table \ref{macol}. 



From Table \ref{macol}, we observe that EXCLIQUE has a globally better performance than MACOL. Indeed, for 9 out of the 12 instances, EXCLIQUE is able to reach a larger best lower bound than MACOL while the reverse is true only for two cases. When it comes to the the average results, the differences between the two approaches are less pronounced. As to the the computing times, it is clear that EXCLIQUE dominates largely MACOL, requiring at most one third of the time required by MACOL. Finally, it is interesting to observe that even if MACOL is inferior to EXCLIQUE, its bounds are still much better than the previous bounds that were obtained with other (less powerful) coloring algorithms.  

\renewcommand{\baselinestretch}{0.7}\large\normalsize
\begin{table}\centering
\begin{small}
\caption{Comparison of EXCLIQUE with the graph coloring approach}\label{macol}
\label{OBO}
\begin{tabular}{p{2.5cm}p{1.0cm}p{1.0cm}p{1.5cm}p{1.5cm}p{0.1cm}p{1.0cm}p{1.5cm}p{1.5cm}p{0cm}}
\hline
\leftline{$Instance$} & \centering{$LB^{*}$} &  \multicolumn{3}{c}{EXCLIQUE} & & \multicolumn{3}{c}{MACOL} & \\
\cline{3-5}\cline{7-9}
& & \centering{$LB_{best}$} & \centering{$Avg$} &  \centering{$T[second]$} & &  \centering{$LB_{best}$} & \centering{$Avg$} &  \centering{$T[second]$} & \\
\hline
\centering{DSJC125.1} & \centering{238}  & \centering{\textbf{246}} &  \centering{244.10} & \centering{80} & & \centering{245} & \centering{244.05} & \centering{285} &  \\
\centering{DSJC125.5} & \centering{504}  & \centering{\textbf{536}} &  \centering{522.40} & \centering{35} & & \centering{534} & \centering{529.55} & \centering{192} &  \\
\centering{DSJC125.9} & \centering{1621}  & \centering{\textbf{1664}} &  \centering{1592.50} & \centering{47} & & \centering{1637} & \centering{1634.25} & \centering{132} &  \\
\centering{DSJC250.1} & \centering{537}  & \centering{567} &  \centering{561.95} & \centering{46} & & \centering{567} & \centering{566.60} & \centering{536} &  \\
\centering{DSJC250.5} & \centering{1150}  & \centering{\textbf{1270}} &  \centering{1258.80} & \centering{37} & & \centering{1266} & \centering{1258.35} & \centering{419} &  \\
\centering{DSJC250.9} & \centering{3972}  & \centering{\textbf{4179}} &  \centering{4082.40} & \centering{158} & & \centering{4062} & \centering{4056.50} & \centering{285} &  \\
\centering{DSJC500.1} & \centering{1163}  & \centering{1250} &  \centering{1246.55} & \centering{1269} & & \centering{\textbf{1257}} & \centering{1254.80} & \centering{1276} &  \\
\centering{DSJC500.5} & \centering{2616}  & \centering{\textbf{2921}} &  \centering{2902.60} & \centering{60} & & \centering{2916} & \centering{2910.75} & \centering{948} &  \\
\centering{DSJC500.9} & \centering{10074}  & \centering{\textbf{10881}} &  \centering{10734.50} & \centering{276} & & \centering{10830} & \centering{10818.20} & \centering{760} &  \\
\centering{DSJC1000.1} & \centering{2499}  & \centering{2762} &  \centering{2758.55} & \centering{5193} & & \centering{\textbf{2775}} & \centering{2719.15} & \centering{13922} &  \\
\centering{DSJC1000.5} & \centering{5787}  & \centering{\textbf{6708}} &  \centering{6665.90} & \centering{155} & & \centering{6545} & \centering{6536.60} & \centering{15119} &  \\
\centering{DSJC1000.9} & \centering{23863}  & \centering{\textbf{26557}} &  \centering{26300.25} & \centering{2741} & & \centering{25879} & \centering{25782.30} & \centering{11258} &  \\
\hline
\end{tabular}
\end{small}
\end{table}

\renewcommand{\baselinestretch}{1.0}\large\normalsize


\section{Analysis of the effect of the disjoint clique removal}

Different methods can be applied to extract cliques from the graph, one simple and conventional method (denoted by OBOCLIQUE) is to extract at each time exactly one maximum clique until the graph becomes empty. Compared to this basic method, our EXCLIQUE algorithm, which uses a heuristic method to extract at each iteration as many disjoint cliques as possible of the largest possible size, is able to identify more large cliques contained in the final decomposition, thus leading to a larger chromatic sum.

We present in this section computational evidences to show the advantage of our disjoint clique removal approach (EXCLIQUE) over the one-by-one clique removal approach (OBOCLIQUE). For OBOCLIQUE, we apply repetitively the ATS clique algorithm \cite{WuHao2010a} to extract one largest possible clique until the graph becomes empty. We consider again the 12 DIMACS random DSJC graphs and report in Table \ref{OBO} a detailed comparison between these two methods. For each instance, we run OBOCLIQUE 20 times and the following statistics are provided: the best lower bounds, the averaged lower bound and the average CPU time in seconds. The results of EXCLIQUE are extracted from Table \ref{dimacs}. 

Table \ref{OBO} discloses that EXCLIQUE dominates OBOCLIQUE in terms of solution quality. Indeed, for each of these 12 instances, EXCLIQUE is able to find a larger lower bound than OBOCLIQUE. To get some insights about the performance difference, we show in Table \ref{decomp} the number of cliques of different sizes extracted by both methods on the instance DSJC1000.5. We see clearly that compared to the OBOCLIQUE method, our EXCLIQUE method is able to extract more larger cliques (leading to a larger chromatic sum, thus better lower bound).

It is also interesting to observe that even if OBOCLIQUE can not compete with EXCLIQUE, the results of OBOCLIQUE remain very competitive compared to the current best lower bounds reported in the literature. This observation highlights the interest of the general clique extraction approach for computing the lower bounds for the MSCP when an effective maximum clique like ATS is employed.

Finally, due to the fact that EXCLIQUE may need to call the ATS clique algorithm many times at each extraction step (only one call per extraction for OBOCLIQUE), EXCLIQUE requires naturally more computing times than OBOCLIQUE. 

\renewcommand{\baselinestretch}{0.7}\large\normalsize
\begin{table}\centering
\begin{small}
\caption{Comparison of clique extraction methods: disjoint clique removal (EXCLIQUE) v.s. one-by-one removal (OBOCLIQUE)}
\label{OBO}
\begin{tabular}{p{2.5cm}p{1.0cm}p{1.0cm}p{1.5cm}p{1.5cm}p{0.1cm}p{1.0cm}p{1.5cm}p{1.5cm}p{0cm}}
\hline
\leftline{$Instance$} & \centering{$LB^{*}$} &  \multicolumn{3}{c}{EXCLIQUE} & & \multicolumn{3}{c}{OBOCLIQUE} & \\
\cline{3-5}\cline{7-9}
& & \centering{$LB_{best}$} & \centering{$Avg$} &  \centering{$T[second]$} & &  \centering{$LB_{best}$} & \centering{$Avg$} &  \centering{$T[second]$} & \\
\hline
\centering{DSJC125.1} & \centering{238}  & \centering{\textbf{246}} &  \centering{244.10} & \centering{80} & & \centering{239} & \centering{232.55} & \centering{20} &  \\
\centering{DSJC125.5} & \centering{504}  & \centering{\textbf{536}} &  \centering{522.40} & \centering{35} & & \centering{524} & \centering{509.6} & \centering{19} &  \\
\centering{DSJC125.9} & \centering{1621}  & \centering{\textbf{1664}} &  \centering{1592.50} & \centering{47} & & \centering{1646} & \centering{1586.20} & \centering{32} &  \\
\centering{DSJC250.1} & \centering{537}  & \centering{\textbf{567}} &  \centering{561.95} & \centering{46} & & \centering{544} & \centering{533.20} & \centering{34} &  \\
\centering{DSJC250.5} & \centering{1150}  & \centering{\textbf{1270}} &  \centering{1258.80} & \centering{37} & & \centering{1234} & \centering{1223.45} & \centering{34} &  \\
\centering{DSJC250.9} & \centering{3972}  & \centering{\textbf{4179}} &  \centering{4082.40} & \centering{158} & & \centering{4116} & \centering{4054.05} & \centering{85} &  \\
\centering{DSJC500.1} & \centering{1163}  & \centering{\textbf{1250}} &  \centering{1246.55} & \centering{1269} & & \centering{1184} & \centering{1174.85} & \centering{391} &  \\
\centering{DSJC500.5} & \centering{2616}  & \centering{\textbf{2921}} &  \centering{2902.60} & \centering{60} & & \centering{2868} & \centering{2840.25} & \centering{51} &  \\
\centering{DSJC500.9} & \centering{10074}  & \centering{\textbf{10881}} &  \centering{10734.50} & \centering{276} & & \centering{10827} & \centering{10687.60} & \centering{160} &  \\
\centering{DSJC1000.1} & \centering{2499}  & \centering{\textbf{2762}} &  \centering{2758.55} & \centering{5193} & & \centering{2635} & \centering{2616.10} & \centering{ 922} &  \\
\centering{DSJC1000.5} & \centering{5787}  & \centering{\textbf{6708}} &  \centering{6665.90} & \centering{155} & & \centering{6549} & \centering{6526.65} & \centering{119} &  \\
\centering{DSJC1000.9} & \centering{23863}  & \centering{\textbf{26557}} &  \centering{26300.25} & \centering{2741} & & \centering{26455} & \centering{26168.85} & \centering{320} &  \\
\hline
\end{tabular}
\end{small}
\end{table}
\bibliographystyle{plain}

\begin{thebibliography}{50}


\bibitem{BarNoy1998}
A. Bar-Noy, M. Bellareb, M. M. Halld\'{o}rsson, H. Shachnai, T. Tamir.
On chromatic sums and distributed resource allocation.
Information and Computation 140(2): 183--202, 1998.

\bibitem{BarNoy1998B}
A. Bar-Noy and G. Kortsarz.
Minimum color sum of bipartite graphs.
Journal of Algorithms 28(2): 339--365, 1998.

\bibitem{Bonomo2011}
F. Bonomo, G. Dur\'{a}n, J. Marenco,  M. Valencia-Pabon.
Minimum sum set coloring of trees and line graphs of trees.
Discrete Applied Mathematics 159(5): 288--294,  2011.

\bibitem{Bouziri2010}
H. Bouziri and M. Jouini. A tabu search approach for the sum coloring problem.
Electronic Notes in Discrete Mathematics 36(1): 915--922, 2010.


\bibitem{Douiri2011}
S.M. Douiri and S. Elbernoussi.
New algorithm for the sum coloring problem.
International Journal of Contemporary Mathematical Sciences 6(10): 453--463, 2011.

\bibitem{Douiri2012}
S.M. Douiri and S. Elbernoussi.
A New Ant Colony Optimization Algorithm for the Lower Bound of Sum Coloring Problem. Journal of Mathematical Modelling and Algorithms 11(2), 181--192, 2012.



\bibitem{Garey1979}
M.R. Garey and D.S. Johnson. Computers and intractability: A guide to the theory of NP-completeness. W.H. Freeman and Company, San Francisco, 1979.

\bibitem{Gandhi2004}
R. Gandhi, M.M. Halld\'{o}rsson, G. Kortsarz and H. Shachnai, Improved bounds for sum multicoloring and scheduling dependent jobs with minsum criteria,
In: Proc. of the Second Workshop on Approximation and Online Algorithms, 68-–82, 2005.


\bibitem{Hajiabolhassan2000}
H. Hajiabolhassan, M.L. Mehrabadi, R. Tusserkani.
Minimal coloring and strength of graphs.
Discrete Mathematics 215(1--3): 265--270, 2000.

\bibitem{Helmar2011}
A. Helmar and M. Chiarandini. A local search heuristic for chromatic sum. In L. D. Gaspero, A. Schaerf, T. St\"{u}tzle (Eds.), Proceedings of the 9th Metaheuristics International Conference, MIC, pp 161--170, 2011.

\bibitem{Jansen1998}
K. Jansen. Approximation results for the optimum cost chromatic partition problem.
Journal of Algorithms 34(1): 54--89, 2000.

\bibitem{Jiang1999}
T. Jiang and D. West.
Coloring of trees with minimum sum of colors.
Journal of Graph Theory 32(4): 354--358, 1999.

\bibitem{Johnson1996}
D.S. Johnson and M.A.Trick (Eds). Cliques, Coloring, and Satisfiability: Second DIMACS Implementation Challenge. DIMACS Series in Discrete Mathematics and Theoretical Computer Science. vol 26, AMS, Providence, RI, 1996.

\bibitem{Kroon1996}
L. G. Kroon, A. Sen, H. Deng, A. Roy.
The optimal cost chromatic partition problem for trees and interval graphs.
Lecture Notes in Computer Science 1197: 279--292, 1996.

\bibitem{Kokosinski2007}
Z. Kokosi\'{n}ski and K. Kawarciany.
On sum coloring of graphs with parallel genetic algorithms.
Lecture Notes In Computer Science 4431: 211--219, 2007.

\bibitem{Kubicka2004}
E. Kubicka. The chromatic sum of graphs; history and recent developments. The International
Journal of Mathematical Sciences, 30: 1563–-1573, 2004.

\bibitem{Kubicka1989}
E. Kubicka and A. J. Schwenk. An introduction to chromatic sums.
Proceedings of the 17th Annual ACM Computer Science Conference, pp 39--45, 1989.

\bibitem{Kubika1991}
E. Kubicka, G. Kubicki, D. Kountanis.
Approximation algorithms for the chromatic sum.
Proceedings of the First Great Lakes Computer Science Conference.
Lecture Notes in Computer Science 507: 15--21, 1991.



\bibitem{Li2009}
Y. Li, C. Lucet, A. Moukrim, K. Sghiouer.
Greedy algorithms for minimum sum coloring algorithm.
Proceedings LT2009 Conference, Tunisia, March 2009.

\bibitem{lu2010}
Z. L\"{u} and J.K. Hao.
A memetic algorithm for graph coloring.
European Journal of Operational Research 200(1): 235--244, 2010.

\bibitem{Malafiejski2004}
M. Malafiejski. Sum coloring of graphs.
Graph Colorings, Contemporary Mathematics 352, AMS, 55--65, 2004.



\bibitem{Moukrim}
A. Moukrim, K. Sghiouer, C. Lucet,  Y. Li.
Lower bounds for the minimal sum coloring problem.
Electronic Notes in Discrete Mathematics 36: 663--670, 2010.







\bibitem{Salavatipour2003}
M. R. Salavatipour.
On sum coloring of graphs.
Discrete Applied  Mathematics 127(3): 477--488, 2003.




\bibitem{WuHao2010a}
Q. Wu and J.K. Hao.
Adaptive multistart tabu search for the maximum clique problem. Accepted to Journal of Combinatorial Optimization, Nov. 2011. DOI: 10.1007/s10878-011-9437-8

\bibitem{WuHaoCOR}
Q. Wu and J.K. Hao. Coloring large graphs based on independent set extraction. Computers and Operations Research, 39(2): 283--290, 2012.

\bibitem{WuHao2012}
Q. Wu and J.K. Hao.
An effective heuristic algorithm for sum coloring of graphs. Computers and Operations Research 39(7): 1593–-1600, 2012.

\end{thebibliography}

\renewcommand{\baselinestretch}{1.0}\large\normalsize
\renewcommand{\baselinestretch}{0.7}\large\normalsize
\begin{table}\centering
\begin{small}
\caption{Detailed comparison on DSJC1000.5 of disjoint clique removal (EXCLIQUE) and one-by-one removal (OBOCLIQUE)}
\label{decomp}
\begin{tabular}{p{2.0cm}p{4.5cm}p{0.1cm}p{2.0cm}p{4.5cm}p{0cm}}
\hline
  \multicolumn{2}{c}{EXCLIQUE} & & \multicolumn{2}{c}{OBOCLIQUE} & \\
\cline{1-2}\cline{4-5}
\centering{$|C|$} & \centering{No. of cliques of size $|C|$} &  &  \centering{$|C|$} & \centering{No. of cliques of size $|C|$} & \\
\hline
\centering{15} & \centering{7} &  &  \centering{15} & \centering{7} & \\
\centering{14} & \centering{26} &  &  \centering{14} & \centering{19} & \\
\centering{13} & \centering{12} &  &  \centering{13} & \centering{15} & \\
\centering{12} & \centering{11} &  &  \centering{12} & \centering{10} & \\
\centering{11} & \centering{6} &  &  \centering{11} & \centering{10} & \\
\centering{10} & \centering{5} &  &  \centering{10} & \centering{6} & \\
\centering{9} & \centering{5} &  &  \centering{9} & \centering{5} & \\
\centering{8} & \centering{2} &  &  \centering{8} & \centering{5} & \\
\centering{7} & \centering{4} &  &  \centering{7} & \centering{1} & \\
\centering{6} & \centering{3} &  &  \centering{6} & \centering{3} & \\
\centering{5} & \centering{1} &  &  \centering{5} & \centering{4} & \\
\centering{4} & \centering{2} &  &  \centering{4} & \centering{0} & \\
\centering{3} & \centering{0} &  &  \centering{3} & \centering{2} & \\
\centering{2} & \centering{3} &  &  \centering{2} & \centering{3} & \\
\centering{1} & \centering{1} &  &  \centering{1} & \centering{2} & \\
\hline
\end{tabular}
\end{small}
\end{table}
\renewcommand{\baselinestretch}{1.0}\large\normalsize

\section{Conclusion}
In this paper, we have presented an effective heuristic algorithm based on clique decomposition for computing lower bounds for the minimum sum coloring problem. Basically, the proposed EXCLIQUE algorithm identifies and removes at each extraction iteration as many disjoint cliques as possible of the same size that is as large as possible. Compared to the conventional one-by-one extraction strategy, EXCLIQUE is able to extract more large cliques from the graph such that a higher number of vertices need to be colored with large color numbers, thus leading to a clique decomposition with a large chromatic sum which corresponds to a tighter lower bound for the MSCP.

We have assessed the performance of EXCLIQUE on a set of 62 benchmark graphs (29 from DIMACS and 33 from COLOR02). For the 38 instances with reported lower bounds, our EXCLIQUE algorithm has improved on the current best lower bounds in 14 cases, proved optimality for the first time for 4 instances and attained the previously best bounds for 22 other instances. Only in two cases, our bounds are slightly worse than the best known bounds. We also reported for the first time lower bounds for the remaining 24 instances that were missing in the literature.

Our comparison between EXCLIQUE and the basic one-by-one clique extraction method has showed a clear dominance of EXCLIQUE in terms of solution quality even though the one-by-one extraction method is faster. We have also revisited the clique decomposition approach based on coloring the complementary graph by using a powerful graph coloring algorithm. This experiment demonstrated that it is globally less effective than the proposed EXCLIQUE even though in some cases the graph coloring approach can attain better results.

\section*{Acknowledgment}

This work was partially supported by the Region of ``Pays de la Loire'' (France) within the Radapop (2009-2012) and LigeRO Projects (2009-2013).

\end{document}